\documentclass[english,reprint, aip,jcp,superscriptaddress,showpacs,floatfix]{revtex4-1}
\usepackage{amsmath}
\usepackage{graphicx, subfigure}
\usepackage{makeidx}

\begin{document}

\title{Surface hopping from the perspective of quantum-classical Liouville dynamics}

%% use optional labels to link authors explicitly to addresses:
%% \author[label1,label2]{}
%% \address[label1]{}
%% \address[label2]{}

\author{Raymond Kapral}

\address{Chemical Physics Theory Group, Department of Chemistry, University of Toronto, Toronto, ON, M5S 3H6 Canada}

\begin{abstract}
Fewest-switches surface hopping is studied in the context of quantum-classical Liouville dynamics. Both approaches are mixed quantum-classical theories that provide a way to describe and simulate the nonadiabatic quantum dynamics of many-body systems. Starting from a surface-hopping solution of the quantum-classical Liouville equation, it is shown how fewest-switches dynamics can be obtained by dropping terms that are responsible for decoherence and restricting the nuclear momentum changes that accompany electronic transitions to those events that occur between population states. The analysis provides information on some of the elements that are essential for the construction of accurate and computationally tractable algorithms for nonadiabatic processes.
\end{abstract}

\maketitle

\section{Introduction} \label{sec:intro}

The Born-Oppenheimer approximation~\cite{born27} figures prominently in studies of quantum structure and dynamics. It relies on a scale separation that is controlled  by a small parameter gauged by the ratio $m/M$ of the light $m$ to heavy $M$ masses of different constituents of the system. This approximation forms the basis for most of electronic structure theory and is also used in adiabatic quantum dynamics where nuclei move on single Born-Oppenheimer surfaces. Although the Born-Oppenheimer approximation has wide utility, it does break down and this breakdown signals the fact that quantum nuclear motion can no longer be described as motion on a single electronic state. Nonadiabatic dynamics is important for the description of many excited-state physical processes. Quantum dynamical methods that account for the breakdown of the Born-Oppenheimer approximation must then be used to follow the time evolution of the system.~\cite{tully12} Fewest-switches surface hopping~\cite{tully90} is one of the most widely used schemes for this purpose. More generally, the basic elements of the fewest-switches algorithm often enter into molecular dynamics methods that involve quantum transitions.~\cite{chernyak11}

In fewest-switches surface hopping the nuclei are assumed follow stochastic trajectories $X_t=(R_t,P_t)$, with $R=(R_1,R_2, \dots,R_N)$ and $P=(P_1,P_2, \dots, P_N)$ denoting the $N$ nuclear positions and momenta, respectively~\cite{footnote:nuclear-notation}. Trajectory evolution takes place on single adiabatic surfaces with stochastic ``hops" to other surfaces that occur with probabilities that are constructed to lead to the fewest number of hops consistent with the electronic populations. While this method has known shortcomings it is simple to use and often yields reasonable results.

More specifically, the equations for the electronic density matrix elements governing the dynamics are as follows: The diagonal density matrix elements satisfy
\begin{equation} \label{eq:sub-den-SH1}
\frac{d}{dt}{\rho}^{\nu \nu}(X_t,t) =-2  \Re \left(\frac{P_t}{M} \cdot  d_{\nu \alpha'}(R_t) \rho^{*\nu \alpha' }(X_t,t)\right),
\end{equation}
while the off-diagonal elements evolve by
\begin{eqnarray} \label{eq:sub-den-SH2}
&&\frac{d}{dt} \rho^{\nu \alpha'}(X_t,t)= -i \omega_{\nu \alpha'}(R_t)\rho^{\nu \alpha'}(X_t,t)\\
 && \qquad  -\frac{P_t}{M} \cdot  d_{\nu \beta}(R_t) \rho^{\beta \alpha'}(X_t,t) -\frac{P_t}{M} \cdot d^*_{\alpha' \beta}(R_t) \rho^{\nu \beta}(X_t,t).\nonumber
\end{eqnarray}
In this equation $d_{\alpha \beta}(R)$ is the nonadiabatic coupling matrix element, $d_{\alpha \beta}(R)= \langle \alpha; R | \nabla_R|\beta; R \rangle$ and $|\alpha; R \rangle$, denotes the $\alpha$ adiabatic eigenstate. The summation convention was used above and will be used throughout the paper except where summations are written in full for clarity.

Transitions between adiabatic states occur probabilistically with a transition rate selected so that the fraction of trajectories in a given adiabatic state corresponds to the electronic population of that state. Energy is conserved along the stochastic trajectories and to ensure that this is the case whenever a nonadiabatic transition causes the system to change its state the nuclear momenta are adjusted to compensate for the energy change in the quantum transition. For example, if a transition from state $\alpha$ to state $\beta$ occurs the momenta of the nuclei along the direction of the nonadiabatic coupling vector are adjusted by $P \to P+\Delta P^{FS}_{\alpha \beta}$, with
\begin{eqnarray} \label{eq:tully-mom-jump}
\Delta P^{FS}_{\alpha \beta}&=& \hat{d}_{\alpha \beta}\;{\rm sgn} (P\cdot \hat{d}_{\alpha \beta}) \sqrt{(P \cdot
\hat{d}_{\alpha \beta})^2 + 2\Delta E_{\alpha \beta} M}\nonumber \\
&&- \hat{d}_{\alpha \beta}(P \cdot \hat{d}_{\alpha \beta}),
\end{eqnarray}
 to conserve energy. Here the energy gap is $\Delta E_{\alpha \beta}=E_\alpha -E_\beta$. For upward transitions it may happen that there is insufficient energy in the nuclear degrees of freedom to insure energy conservation. In this case the transition rule needs to be modified, usually by setting the transition probability to zero.

This algorithm captures many of the important physical features of nonadiabatic dynamics and is easy to implement in computations, thus justifying its widespread use. It is not without defects. Since there is no mechanism for the decay of the off-diagonal density matrix elements, it cannot describe the effects of decoherence on nonadiabatic processes. A considerable amount of effort has been devoted to modification of fewest-switches surface hopping to introduce decoherence into the scheme.~\cite{hammesschiffer94,bittner95,bittner97,schwartz96,truhlar04,subotnik11,subotniki11a,subotnik11c,subotnik11d,prezhdo12}

The aim of this article is to determine the conditions under which quantum-classical Liouville dynamics~\cite{footnote:annrevpchem} can be reduced to fewest-switches surface hopping. The quantum-classical Liouville equation provides a basis for the derivation of various quantum-classical methods.~\cite{kapral15} While solutions to this equation can be obtained by a variety of methods, solutions may also be obtained by a surface-hopping algorithm~\cite{mackernan02,sergi03b,mackernan08}, and it is in the context of the approximations to this surface-hopping dynamics that we shall consider fewest-switches surface hopping. In particular, it will be shown that by dropping terms that account for the effects of decoherence and modifying how nonadiabatic transitions and the nuclear momentum changes that accompany them are treated, one can arrive fewest-switches surface hopping.

The main text begins in Sec.~\ref{sec:qcl} with a brief outline and critical discussion of the features of the quantum-classical Liouville equation in the adiabatic basis and its solution by a surface-hopping algorithm. This sets the stage for the analysis in Sec.~\ref{sec:surf-hop} that allows one to see in some detail the approximations to the dynamics that lead to fewest-switches surface hopping. The last section of the paper discusses how the results of this study may provide ingredients for the construction of new surface-hopping algorithms.

%%%%%%%%%%%%%%%%%%%%%%%%%%%%%%%%%%%%%%%%%
\section{Quantum-Classical Liouville Dynamics in the Adiabatic Basis}\label{sec:qcl}

Since surface-hopping methods are often formulated in the adiabatic basis, it is instructive to discuss the dynamical picture that emerges when the quantum-classical Liouville equation (QCLE) is expressed in this basis. The partially Wigner transformed Hamiltonian, $\hat{H}_W$, for the system can be written as the sum of the nuclear kinetic energy, $P^2/2M$, and the remainder of the electronic, nuclear and coupling terms contained in the operator $\hat{h}(R)$: $\hat{H}_W=P^2/2M +\hat{h}(R)$. The adiabatic energies, $E_\alpha (R)$, and the adiabatic states, $ | \alpha; R \rangle$ are determined from the solution of the eigenvalue problem, $\hat{h}(R)| \alpha; R \rangle = E_\alpha (R) | \alpha; R \rangle$, and depend parametrically on the coordinates of the nuclei.  Adopting an Eulerian description where the dynamics is viewed at a fixed nuclear phase space point $X=(R,P)$, the QCLE for the density matrix elements, $ \langle \alpha; R | \hat{\rho}_W (X, t) |\alpha';R \rangle=\rho_W^{\alpha \alpha'} (X, t)$ in the adiabatic basis is~\cite{kapral99}
\begin{eqnarray}\label{eq:adiabatic_qcle}
&&   \frac{\partial}{\partial t} \rho^{\alpha \alpha'}_W (X, t) = - (i\omega_{\alpha \alpha'} + iL_{\alpha \alpha'})\rho^{\alpha \alpha'}_W (X, t) \\
   &&\qquad+ {\mathcal J}_{\alpha \alpha', \beta \beta'} \rho^{\beta \beta'}_W (X, t) \equiv - i  {\cal L}_{\alpha \alpha', \beta \beta'} \rho^{\beta \beta'}_W (X, t). \nonumber
\end{eqnarray}
The frequency is $\omega_{\alpha \alpha'}(R)=(E_{\alpha}-E_{\alpha'})/\hbar \equiv \Delta E_{\alpha\alpha'}(R)/\hbar$  and the classical Liouville operator $iL_{\alpha \alpha'}$ is defined by
\begin{equation}\label{eq:adiabatic_superoperator}
iL_{\alpha \alpha'}=\frac{P}{M}\cdot{\partial \over \partial R} +
\frac{1}{2} \left( F_{\alpha} + F_{\alpha'} \right)
\cdot{\partial  \over \partial P},
\end{equation}
where the Hellmann-Feynman forces are $F_{\alpha}(R)= -{\partial E_\alpha(R)}/{\partial R}$. The operator,
\begin{eqnarray}\label{eq:jdef}
   {\mathcal J}_{\alpha \alpha',\beta \beta'} & = & - d_{\alpha \beta} \cdot \left(\frac{P}{M}
+\frac{1}{2} \Delta E_{\alpha\beta}{\partial  \over \partial P}\right)\delta_{\alpha' \beta'} \nonumber \\
   & - & d^*_{\alpha' \beta'}\cdot \left(\frac{P}{M}+\frac{1}{2} \Delta E_{\alpha' \beta'}
{\partial  \over \partial P}\right) \delta_{\alpha \beta},
\end{eqnarray}
couples the dynamics on the individual and mean adiabatic surfaces. The last line of Eq.~(\ref{eq:adiabatic_qcle}) defines the QCL operator, $ i  {\cal L}_{\alpha \alpha', \beta \beta'}$.

A few features of this equation are worth noting. The classical evolution operators $iL_{\alpha \alpha'}$ describe adiabatic evolution on either single ($\alpha =\alpha'$) surfaces or on the mean of two surfaces when $\alpha \ne \alpha'$. No approximation is made to obtain such evolution on the mean of two surfaces for off-diagonal elements; it follows naturally from the representation of the QCLE in the adiabatic basis. The coupling term ${\mathcal J}_{\alpha \alpha',\beta \beta'}$ not only involves nonadiabatic coupling matrix elements, $d_{\alpha \beta}(R)$, but also derivatives with respect to the nuclear momenta. This term accounts for part of the influence of the nonadiabatic quantum electronic dynamics on the nuclei. This important coupling adds complexity to the equation of motion and its exact treatment precludes a simple description of the nuclear evolution.

One may attempt to solve this equation by any convenient method and considerable effort and schemes have been devised with the aim of obtaining accurate yet computationally tractable solutions.~\cite{donoso98,wan00a,santer01,wan02,horenko02,kim-map08,nassimi10,kelly12,hsieh12,kelly-markland13,hsieh13,rhee14}
Since the goal of this paper is to explore connections to fewest-switches surface hopping (FSSH), we consider approximate solutions that are based on surface-hopping trajectories. It is useful to observe that while the QCLE conserves energy, nothing is implied about conservation of energy in any single trajectory that might be used in solutions to this equation.

\subsection*{Surface-hopping solution of the QCLE}

The basis for the surface-hopping solution was described some time ago~\cite{kapral99} and the details of the algorithms and their applications to various problems have been discussed previously~\cite{mackernan02,sergi03b,mackernan08}. Nevertheless, it is useful to present a brief account of this solution scheme in order to contrast it with FSSH in the next section, and to point to some of its features that are often overlooked. In general terms the surface-hopping method is a stochastic algorithm for the solution of the QCLE that relies on Monte Carlo sampling of diagonal and off-diagonal electronic states and accounts for nuclear momentum changes when transitions occur. In its usual implementation only one basic approximation is made: the momentum-jump approximation~\cite{kapral99,0chap-kapral02,footnote:annrevpchem,kapral15}. This approximation, outlined below, replaces the infinitesimal nuclear momentum changes contained in the ${\mathcal J}_{\alpha \alpha',\beta \beta'}$ coupling term by finite momentum changes. The approximation both makes the dynamics much more tractable computationally and provides a link to other surface-hopping schemes. For example, if instead of using the momentum-jump approximation momentum derivatives are approximated by finite differences, an exponentially increasing branching tree of trajectories results that quickly makes computation  intractable.~\cite{nielsen00b} Other simulation schemes cited above that are not based on surface hopping do not make the momentum-jump approximation.

The momentum-jump approximation begins by rewriting the operators that appear in ${\mathcal J}$ as
\begin{equation}\label{eq:Jform}
d_{\alpha \beta} \cdot \left(\frac{P}{M} +\frac{1}{2}E_{\alpha\beta}{\partial  \over \partial P}\right)  = \frac{P}{M} \cdot d_{\alpha \beta}
 \left(1+ M \Delta E_{\alpha \beta} \frac{\partial}{\partial {\mathcal Y}_{\alpha \beta}} \right),
\end{equation}
where ${\mathcal Y}_{\alpha \beta}=(P \cdot \hat{d}_{\alpha \beta})^2$.
This form shows that the momentum changes can be expressed in terms of an $R$-dependent prefactor ($\Delta E_{\alpha \beta}(R)$) multiplying a derivative with respect to the square of the momentum along $\hat{d}_{\alpha \beta}$. The momentum-jump approximation replaces the factor in parentheses on the right side by an exponential operator with the same leading terms,
\begin{equation}
 \left(1+ M \Delta E_{\alpha \beta} \frac{\partial}{\partial {\mathcal Y}_{\alpha \beta}} \right) \approx e^{M \Delta E_{\alpha \beta} \frac{\partial}{\partial {\mathcal Y}_{\alpha \beta}}} \equiv \hat{j}_{\alpha \beta}.
\end{equation}
When the momentum-jump operator $\hat{j}_{\alpha \beta}$ acts on any function $f(P)$ it yields $\hat{j}_{\alpha \beta}f(P)=f(P+\Delta P_{\alpha \beta})$ where
\begin{eqnarray}
\Delta P_{\alpha \beta}&=& \hat{d}_{\alpha \beta}\;{\rm sgn} (P\cdot \hat{d}_{\alpha \beta}) \sqrt{(P \cdot
\hat{d}_{\alpha \beta})^2 + \Delta E_{\alpha \beta} M}\nonumber \\
&&- \hat{d}_{\alpha \beta}(P \cdot \hat{d}_{\alpha \beta}).
\end{eqnarray}
Apart from a factor of two multiplying $\Delta E_{\alpha \beta}$, this expression for the momentum adjustment is identical to that in Eq.~(\ref{eq:tully-mom-jump}) for the FSSH algorithm. This factor-of-two difference has its origin in the transitions to off-diagonal states (coherences) that take place in QCL dynamics. These results can then be used to write the momentum-jump approximation to ${\mathcal J}$:
\begin{eqnarray}\label{eq:jumpJ}
   {\mathcal J}_{\alpha \alpha',\beta \beta'}(X) &\approx& - \frac{P}{M} \cdot d_{\alpha \beta}(R)\hat{j}_{\alpha\beta}(X)\delta_{\alpha' \beta'}\nonumber \\
  && -\frac{P}{M} \cdot d^*_{\alpha' \beta'}(R)\hat{j}_{\alpha' \beta'}(X)\delta_{\alpha \beta}.
\end{eqnarray}
This form will be used henceforth in the surface-hopping solution of the QCLE.

The surface-hopping solution proceeds as follows: Since the QCL operator commutes with itself the solution of the QCLE can be written exactly as
\begin{equation}
%&&
\rho^{\alpha \alpha'}_W(X,t)=
%\\
%&& \quad
\prod_{j=1}^N \Big( e^{-i {\mathcal L} \Delta t_j}
\Big)_{\alpha_{j-1}\alpha_{j-1}', \alpha_{j}\alpha_{j}'}\rho^{\alpha_N \alpha_N'}_W(X,0),
%\nonumber
\label{eq:formal}
\end{equation}
where the time interval $t$ was divided into $N$ segments of lengths $\Delta t_j=t_j-t_{j-1}= \Delta t$ and $\alpha_0=\alpha$ and $\alpha_0'=\alpha'$. If $\Delta t$ is chosen to be sufficiently small, In each short time segment we may write
\begin{eqnarray}\label{eq:short-time}
&&\Big( e^{-i {\mathcal L} \Delta t}
\Big)_{\alpha_{j-1}\alpha_{j-1}', \alpha_{j}\alpha_{j}'}
\approx {\mathcal W}_{\alpha_{j-1}\alpha_{j-1}'}(\Delta t)e^{-iL_{\alpha_{j-1}\alpha_{j-1}'}\Delta t} \nonumber \\
&&\quad \qquad \times \left(\delta_{\alpha_{j-1} \alpha_{j}} \delta_{\alpha_{j-1}' \alpha_{j}'} + \Delta t {\mathcal J}_{\alpha_{j-1}\alpha_{j-1}', \alpha_{j}\alpha_{j}'}, \right).
\end{eqnarray}
where the phase factor ${\mathcal W}_{\alpha \beta}$ is defined as
\begin{equation}\label{eq:wphase1}
   \mathcal {W}_{\alpha \beta}(t_1,t_2)=  e^{-i\int_{t_1}^{t_2} d\tau \; \omega_{\alpha \beta}(R^{\alpha \beta}_\tau)},
\end{equation}
and the superscript $\alpha \beta$ on $R_\tau$ indicates that it is propagated classically on the mean of the $\alpha$ and $\beta$ surfaces.

The solution for the density matrix then follows from substitution of these short time propagators into Eq.~(\ref{eq:formal}). In principle one just has to carry out the matrix multiplications and actions of the classical evolution and jump operators to find the solution. A better and more computationally tractable way to do this is to sample the electronic states in the matrix multiplications and the actions of the nonadiabatic transitions by a Monte Carlo procedure. In the simplest Monte Carlo scheme the quantum states are uniformly sampled from the allowed set of states and the actions of the nonadiabatic coupling operators are sampled based on a weight function that reflects the magnitude of the nonadiabatic coupling. A simple choice to determine if a transition occurs is
\begin{equation}
\pi =\Big|\frac{P}{M} \cdot d_{\alpha \beta}\Big| \Delta t /\Big( 1 +\Big|\frac{P}{M} \cdot d_{\alpha \beta}\Big| \Delta t \Big),
\end{equation}
but other probability factors have been suggested~\cite{sergi13,hanna-filter16}. If no transition occurs, then a weight $1/(1 - \pi)$ is included to account for this failure. If a transition does occur, a weight $1/\pi$ is applied and the nuclear momenta are adjusted by the momentum-jump operator so that energy is conserved. Note that because of the use of the momentum-jump approximation it may happen that nuclear degrees of freedom do not have sufficient energy for this process to take place. Then the argument of the square root in the expression for $\Delta P_{\alpha \beta}$ will be negative and the expression cannot be used. In this circumstance the transition is not allowed and the evolution continues on the current adiabatic surface.

\begin{figure}[htbp]
\begin{center}
\includegraphics[width=0.9\columnwidth]{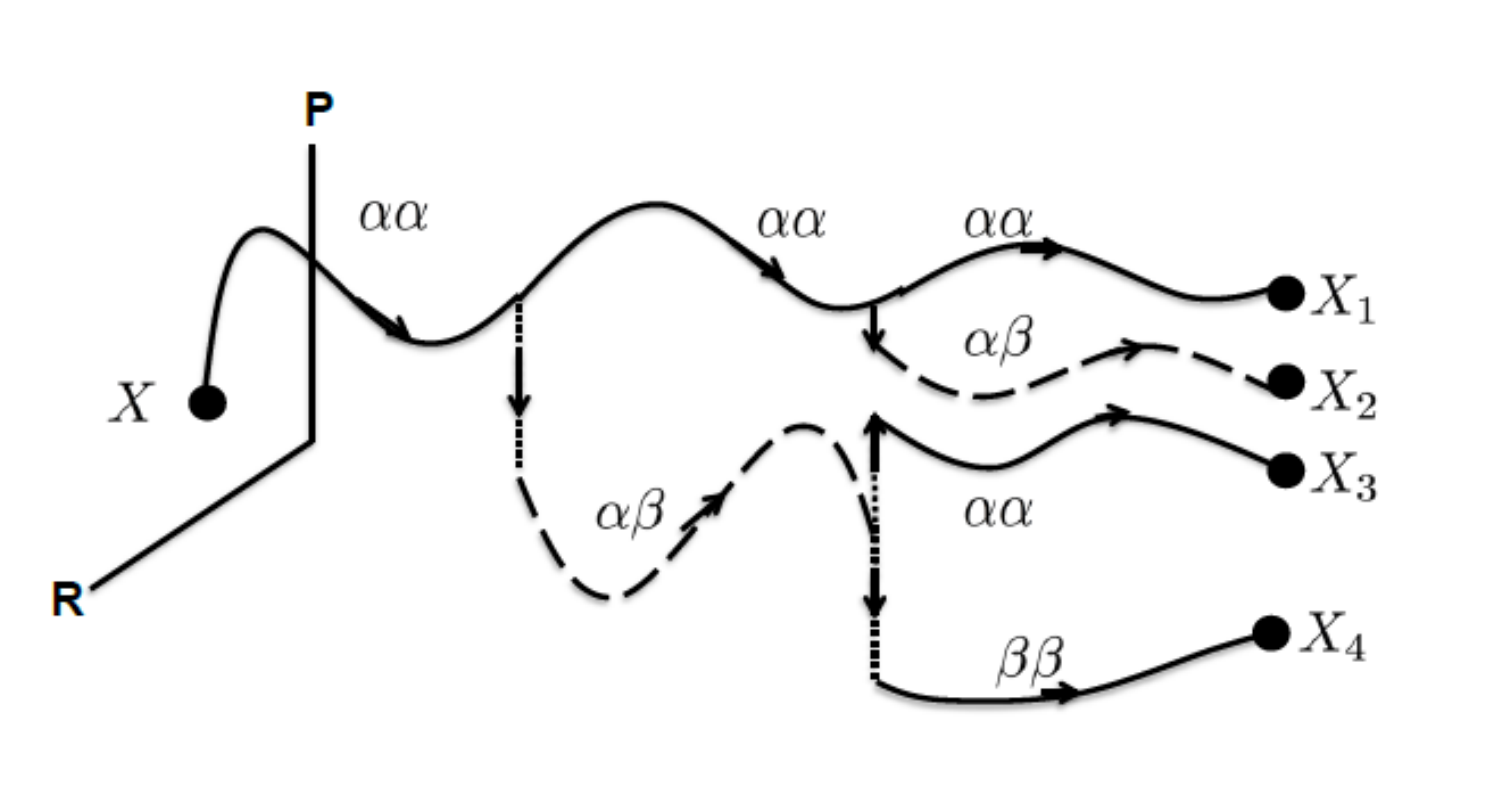}
%\vspace{20pt}
\caption{\label{fig:QCLE-traj} Stochastic trajectories that contribute to the diagonal density matrix $\rho^{\alpha \alpha}_W(X,t)$. The density is computed at a fixed phase space point $X$ in this Eulerian representation. The different trajectories indicate possible sequences of transitions that may occur. In this short set of segments a variety of density matrix elements at different phase points all contribute to the $\alpha \alpha$ density at phase point $X$ at time $t$. The solid lines denote propagation on single adiabatic surfaces while the dashed lines signify propagation on the mean of two adiabatic surfaces; a phase factor is attached to these segments. The vertical dotted lines indicate nonadiabatic surface-hopping transitions accompanied by momentum shifts. In the stochastic algorithm described in the text each trajectory also carries a Monte Carlo weight.}
\end{center}
\end{figure}

From this description one sees that the surface-hopping trajectories are a consequence of the momentum-jump approximation and the Monte Carlo sampling method used to construct the solution. The scheme does not make any anzatz on the nature of the stochastic trajectories that underlie the dynamics nor is any special physical significance attached to the probabilities with which the stochastic hops are carried out. Figure~\ref{fig:QCLE-traj} shows an example of some of the trajectories that contribute to the diagonal ($\alpha \alpha$) density matrix element at phase point $X$ at time $t$.

In practice computations of populations or coherences are carried out somewhat differently by making use of the expression for the average value of an operator, $\hat{{\mathcal O}}(X)$,  given in the adiabatic basis by
\begin{eqnarray}
\overline{{\mathcal O}}(t)&=& \sum_{\alpha \alpha'}\int dX \;{\mathcal O}_{\alpha \alpha'}(X)  \rho^{\alpha' \alpha}_W(X,t)\nonumber \\
&=&\sum_{\alpha \alpha'}\int dX \; {\mathcal O}_{\alpha \alpha'}(X,t) \rho^{\alpha' \alpha}_W(X,0).
\end{eqnarray}
The second line of this equation expresses the expectation value in a computationally more convenient form that involves sampling over the initial density matrix. The time evolution of the operator also satisfies a QCLE but with forward time propagation.~\cite{kapral99} For example, to compute the population in state $\nu$, $\rho_s^{\nu}(t)$, select ${\mathcal O}^\nu_{\alpha \alpha'}(X)=\delta_{\alpha \nu}\delta_{\alpha' \nu}$ so that
\begin{eqnarray}
\rho_s^{\nu}(t)&=&\int dX \;  \rho^{\nu \nu}_W(X,t)=\sum_{\alpha \alpha'}\int dX \;\delta_{\alpha \nu}\delta_{\alpha' \nu}  \rho^{\alpha' \alpha}_W(X,t) \nonumber \\
&=&\sum_{\alpha \alpha'}\int dX \;  \rho^{\alpha \alpha'}_W(X,0) \Big( e^{i {\mathcal L}t}\Big)_{\alpha' \alpha, \nu \nu}.
\end{eqnarray}

From this expression one can see that the time evolved operator will contain all of the reweighting factors needed to obtain the correct population from the average over the ensemble of stochastic trajectories. The population is not obtained by simply determining the fraction of trajectories in state $\nu$ at time $t$. Instead, each trajectory carries a set of weights that give the correct weighting of that trajectory to its contribution in the ensemble. In this way all the correlations in the ensemble are taken into account. This feature is both its most important attribute and the source of its primary difficulty: Monte Carlo weights can accumulate over long trajectories leading to instabilities requiring increasing numbers of trajectories to obtain converged results. The difficulties can partially eliminated by filtering, and filtering methods have been suggested and used in calculations.~\cite{hanna05,mackernan08,sergi13} The method has been shown to give accurate solutions, although the number of trajectories required to obtain the results is considerably larger than for FSSH.

\section{Approximations to yield fewest-switches surface hopping} \label{sec:surf-hop}

Fewest-switches surface hopping assumes that between nonadiabatic hops the nuclear degrees of freedom evolve classically on single adiabatic surfaces governed by Hellmann-Feynman forces.  Consequently, it is convenient to view QCL dynamics in a Lagrangian frame of reference that moves with the nuclear phase space flow along a single adiabatic surface. Letting $\nu$ be the label of the chosen adiabatic surface, the evolution of the nuclear phase space coordinates is given by $X^\nu_t=\exp(iL_\nu t) X$, and they satisfy the usual equations of motion,
\begin{equation}
\dot{R}^\nu_t=\frac{P^\nu_t}{M}, \quad \dot{P}^\nu_t=-\frac{\partial }{\partial R^\nu_t} E_\nu(R^\nu_t).
\end{equation}
Since we can write $ \rho^{\alpha \alpha'}_W (X^\nu_t, t)=\exp(iL_\nu t) \rho^{\alpha \alpha'}_W (X, t)$, the QCLE  in this frame of reference takes the form
\begin{eqnarray}\label{eq:LE-1surf}
 \frac{d}{dt} \rho^{\alpha \alpha'}_W (X^\nu_t, t) &=&- i {\mathcal L}^{(\nu)}_{\alpha \alpha', \beta \beta'}(X^\nu_t)\rho^{\beta \beta'}_W (X^\nu_t,t),
\end{eqnarray}
where the material derivative  specifies the rate of change in this frame and the evolution operator is given by
\begin{eqnarray}\label{eq:L-off-nu}
 i {\mathcal L}^{(\nu)}_{\alpha \alpha', \beta \beta'}(X) &=& i\omega_{\alpha \alpha'}(R) \delta_{\alpha \beta} \delta_{\alpha' \beta'}\nonumber\\
&& +  \frac{1}{2}(\Delta F_{\alpha  \nu}(R)+\Delta F_{\alpha' \nu}(R))\cdot
\frac{\partial}{\partial P} \delta_{\alpha \beta} \delta_{\alpha' \beta'} \nonumber \\
&& - {\mathcal J} _{\alpha \alpha', \beta \beta'}(X) .
\end{eqnarray}

In order to appreciate the content of Eq.~(\ref{eq:LE-1surf}) it is convenient to define formally ``decoherence" factors as
\begin{equation}\label{eq:decoh}
\gamma_{\alpha \alpha'}^{(\nu)}(X) = \frac{1}{2}(\Delta F_{\alpha  \nu}(R)+\Delta F_{\alpha' \nu}(R))\cdot \frac{1}{\rho_W^{\alpha \alpha'}(X)}
\frac{\partial \rho_W^{\alpha \alpha'}(X)}{\partial P}.
\end{equation}
Using this definition the equation of motion takes the form,
\begin{eqnarray}
 \frac{d}{d t} \rho^{\alpha \alpha'}_W (X^\nu_t, t) &=&
 \Big(-i\omega_{\alpha \alpha'}(R^\nu_t) -\gamma_{\alpha \alpha'}^{(\nu)}(X^\nu_t)\Big) \rho^{\alpha \alpha'}_W (X^\nu_t,t) \nonumber \\
 &&+ {\mathcal J} _{\alpha \alpha', \beta \beta'}(X^\nu_t)\rho^{\beta \beta'}_W (X^\nu_t,t).
\end{eqnarray}
The appearance of the decoherence factor in this equation is a consequence of viewing the dynamics on single adiabatic surfaces. It appears in both the equations for the off-diagonal ($\alpha \ne \alpha'$ ) and diagonal ($\alpha =\alpha'$ with $\alpha \ne \nu$) density matrix elements. Note also that if $\alpha' \ne \alpha =\nu$ the decoherence factor takes the simpler form,
\begin{equation}\label{eq:decoh-nu}
\gamma_{\nu \alpha'}^{(\nu)}(X) = \frac{1}{2}\Delta F_{\alpha' \nu}(R)\cdot \frac{1}{\rho_W^{\nu \alpha'}(X)}
\frac{\partial \rho_W^{\nu \alpha'}(X)}{\partial P}.
\end{equation}
This decoherence factor appeared earlier in a study of surface hopping in the context of the QCLE by Subotnik, Ouyang and Landry~\cite{subotnik11c}. While formally exact it is not easily computed but its approximate evaluation has been discussed in this paper. It can form the basis for approximate methods for incorporating decoherence effects in simple surface-hopping schemes.

Writing these equations more explicitly, the equation of motion for the diagonal element of the density matrix for state $\nu$ is,
\begin{equation}\label{eq:diag-nu2}
\frac{d}{dt}{\rho}_W^{\nu \nu}(X^\nu_t,t) =-2\Re \left(\frac{P^\nu_t}{M} \cdot  d_{\nu \alpha'}(R^\nu_t) \hat{j}_{\nu \alpha'}\rho_W^{ * \nu \alpha'}(X^\nu_t,t)\right),
\end{equation}
while the equation for the off-diagonal elements is
\begin{eqnarray}\label{eq:off-diag-nu2}
 \frac{d}{d t} \rho^{\nu \alpha'}_W (X^\nu_t, t) &=&
 \Big(-i\omega_{\nu \alpha'}(R^\nu_t) -\gamma_{\nu \alpha'}^{(\nu)}(X^\nu_t)\Big) \rho^{\nu \alpha'}_W (X^\nu_t,t) \nonumber \\
 &&-\frac{P^\nu_t}{M} \cdot  d_{\nu \beta}(R^\nu_t) \hat{j}_{\nu \beta}\rho^{\beta \alpha'}_W (X^\nu_t,t)\nonumber \\
&& -\frac{P^\nu_t}{M} \cdot  d_{\alpha' \beta}(R^\nu_t) \hat{j}_{\alpha' \beta}\rho^{\nu \beta}_W (X^\nu_t,t).
\end{eqnarray}

These equations are equivalent to the original QCLE (with the momentum-jump approximation), but simply viewed in a different frame. The decoherence term has the form of a classical operator that acts on the nuclear momenta and depends on the difference between two Hellmann-Feynman forces corresponding to two different adiabatic surfaces.  As discussed earlier, coherence is created in the QCLE by transition events that take the system to off-diagonal density matrix elements, and coherence is destroyed when the system returns to a diagonal population state. The decoherence factors that appear in the  above equations are another representation of these processes.

\subsection*{Approximations to these equations}

In FSSH the classical dynamics follows stochastic trajectories comprising evolution on single adiabatic surfaces interrupted by transitions to other
adiabatic surfaces. These transitions are accompanied by momentum adjustments to conserve energy. There are no transitions to off-diagonal density matrix elements. Consequently, to make connection to FSSH, nonadiabatic transitions must be restricted to those events that connect diagonal density matrix elements.

We are now in a position to make approximations to the evolution equations (\ref{eq:diag-nu2}) and (\ref{eq:off-diag-nu2}) that will bring us close to the equations that underlie FSSH. In particular, two approximations connected with decoherence and momentum adjustments need to be made concurrently, and a third approximation concerns the probabilities with which nonadiabatic transitions occur.

 (1) Since decoherence is not taken into account in FSSH, we drop the decoherence factors, $\gamma_{\alpha \alpha'}^{(\nu)}$ in Eq.~(\ref{eq:off-diag-nu2}) to get, for all $\alpha$ and $\alpha'$,
\begin{eqnarray}\label{eq:rho-nu2-nodcoh}
 \frac{d}{d t} \rho^{\alpha \alpha'}_W (X^\nu_t, t)&=&
-i\omega_{\alpha \alpha'}(R^\nu_t)  \rho^{\alpha \alpha'}_W (X^\nu_t,t) \nonumber\\
&&+{\mathcal J}_{\alpha \alpha',\beta \beta'}(X^\nu_t) \rho^{\beta \beta'}_W (X^\nu_t,t),
\end{eqnarray}
where ${\mathcal J}$ is evaluated in the momentum-jump approximation.
We can write this equation more compactly by defining $N^\nu_{\alpha \alpha',\beta \beta'}(t)= -i\omega_{\alpha \alpha'}(R^\nu_t) \delta_{\alpha \beta}\delta_{\alpha' \beta'}+{\mathcal J}_{\alpha \alpha',\beta \beta'}(X^\nu_t)$:
\begin{equation}\label{eq:nodecoh}
 \frac{d}{d t} \rho^{\alpha \alpha'}_W (X^\nu_t, t) = N^\nu_{\alpha \alpha',\beta \beta'}(t) \rho^{\beta \beta'}_W (X^\nu_t,t).
\end{equation}
Now, between nonadiabatic transition events, the evolution of the nuclear degrees of freedom is governed by motion on the currently active single adiabatic surface (the adiabatic state on which propagation is currently taking place -- denoted by $\nu$ here).

(2) In FSSH transitions occur between the active population state and other adiabatic population states. No hops to off-diagonal states, along with their associated momentum jumps, take place. In the context of the QCLE, this means that all momentum-jump operators should be associated solely with transitions involving population states. Jump operators should not be allowed to act when coherences or inactive population states are being propagated.

To see how to implement and appreciate the nature of this approximation it is convenient to rewrite Eq.~(\ref{eq:nodecoh}) as a generalized master equation for the diagonal density matrix elements since this makes the coupling between population states evident. Adopting the procedure used to derive a generalized master equation from the QCLE~\cite{grunwald07}, we denote the diagonal and off-diagonal density matrix elements by $ \rho_d (X^\nu_t, t)$ and $ \rho_o (X^\nu_t, t)$, respectively, and block $N^\nu$ into diagonal, off-diagonal, and coupling components, $N^{\nu d}$, $N^{\nu o}$, $N^{\nu do}$ and $N^{\nu od}$, respectively. Then, formally solving for the off-diagonal density matrix elements and substituting the result into the equation for the diagonal element of the active surface yields~\cite{footnote:off-initial},
\begin{equation}\label{eq:master}
\frac{d}{d t} \rho^{\nu \nu}_W (X^\nu_t, t)= \int_0^t dt' \; {\mathcal M}^\nu_{\nu \beta}(t,t') \rho^{\beta \beta}_W (X^\nu_{t'}, t'),
\end{equation}
where
\begin{equation}\label{eq:mem-kernel}
{\mathcal M}^\nu_{\nu \beta}(t,t')={\mathcal J}^{do}_{\nu, \mu_1 \mu_1'}(X^\nu_t) {\mathcal U}^{\nu o}_{\mu_1 \mu_1', \mu_2 \mu_2'}(t,t')
{\mathcal J}^{od}_{\mu_2 \mu_2', \beta}(X^\nu_{t'}),
\end{equation}
and the simpler notation ${\mathcal J}_{\nu \nu, \mu \mu'}={\mathcal J}^{do}_{\nu,\mu \mu'}$, etc. was used. The propagator for off-diagonal elements is ${\mathcal U}^{\nu o}(t,t')$ and it takes the form of a time-ordered exponential whose power series is
\begin{eqnarray}\label{eq:Uexpansion}
&&\hspace{-0.5cm} {\mathcal U}^{\nu o}_{\mu_1 \mu_1', \mu_2 \mu_2'}(t,t')= \delta_{\mu_1 \mu_2} \delta_{\mu_1' \mu_2'}+ \int_{t'}^t d t_1 \; N^\nu_{\mu_1 \mu_1', \mu_2 \mu_2'}(t_1) \\
&&  + \int_{t'}^t d t_1 \; N^\nu_{\mu_1 \mu_1', \mu_3 \mu_3'}(t_1)  \int_{t'}^{t_1} d t_2 \; N^\nu_{\mu_3 \mu_3', \mu_2 \mu_2'}(t_2)+\cdots \nonumber
\end{eqnarray}
From the definition of $N^\nu$ one can see that ${\mathcal U}^{\nu o}(t,t')$ contains the adiabatic frequencies, nonadiabatic coupling matrix elements and jump operators.~\cite{footnote:non-two-level}

Considering the structure of Eq.~(\ref{eq:mem-kernel}), we see that momentum jump operators at different times appear in the left-most and right-most ${\mathcal J}$ operators, as well as in the off-diagonal propagator. They act on all quantities to their right. Since transitions are only allowed between population states in FSSH we make the approximation that all momentum jump operators are moved through the intervening functions and operators in ${\mathcal M}^\nu$ and are taken to act only on the diagonal density matrix elements $\rho^{\beta \beta}_W (X^\nu_{t'}, t')$ at time $t'$ in Eq.~(\ref{eq:master}). This process will lead to a product of momentum jump operators acting on the populations and this product of operators must be concatenated to obtain the net momentum change.

We compute a few representative terms to show the result of such a concatenation process. Consider the identity operator in the first term in Eq.~(\ref{eq:Uexpansion}). The resulting contribution to the memory kernel is
\begin{eqnarray}
&&\hspace{-0.5cm}{\mathcal M}^{\nu (1)}_{\nu \beta}(t,t')= \\
&& \quad 2\Re \left\{\frac{P^\nu_t}{M} \cdot  d_{\nu \mu_1}(R^\nu_t)  \frac{P^\nu_{t'}}{M} \cdot  d_{ \mu_1 \nu}(R^\nu_{t'})  \right\}\delta_{\nu \beta}
\hat{j}_{\nu \mu_1}\hat{j}_{\mu_1 \nu }\nonumber \\
&& \quad +2\Re \left\{\frac{P^\nu_t}{M} \cdot  d_{\nu \beta}(R^\nu_t)  \frac{P^\nu_{t'}}{M} \cdot  d^*_{ \nu \beta}(R^\nu_{t'}) \right\}
\hat{j}_{\nu \beta}\hat{j}_{\nu \beta}.\nonumber
\end{eqnarray}
The action of two consecutive QCL momentum shifts on some function $f(P)$ can be computed as follows:
\begin{eqnarray}
&&\hat{j}_{\nu \beta}(P) \hat{j}_{\nu \beta}(P) f(P)= \hat{j}_{\nu \beta}(P+\Delta P_{\nu \beta}(P)) f(P+\Delta P_{\nu \beta}(P))\nonumber \\
&& \qquad = f(P+\Delta P_{\nu \beta}(P)+\Delta P_{\nu \beta}(P+\Delta P_{\nu \beta}(P))).
\end{eqnarray}
After some algebra one may show that
\begin{eqnarray}
&&\Delta P_{\nu \beta}(P+\Delta P_{\nu \beta}(P))=-(P+\Delta P_{\nu \beta}(P))\cdot \hat{d}_{\nu \beta} \hat{d}_{\nu \beta} \nonumber\\
&& \qquad+\hat{d}_{\nu \beta}\;{\rm sgn}(P \cdot \hat{d}_{\nu \beta})\sqrt{(P \cdot \hat{d}_{\nu \beta})^2 +2 \Delta E_{\nu \beta}M}.
\end{eqnarray}
Thus, using this result we find that
\begin{equation}
\Delta P_{\nu \beta}(P)+\Delta P_{\nu \beta}(P+\Delta P_{\nu \beta}(P))=\Delta P^{FS}_{\nu \beta}(P),
\end{equation}
and we can write,
\begin{equation}
\hat{j}_{\nu \beta}(P) \hat{j}_{\nu \beta}(P) f(P)=\hat{j}^{FS}_{\nu \beta}(P)f(P)=f(P+\Delta P^{FS}_{\nu \beta}(P)).
\end{equation}
We have used the fewest-switches (FS) superscript on this jump operator to indicate that it produces the same momentum shift as that in FSSH given in Eq.~(\ref{eq:tully-mom-jump}).  Following the same procedure we find that
\begin{equation}
\hat{j}_{\nu \mu_1}(P) \hat{j}_{\mu_1 \nu }(P) f(P)=\hat{j}^{FS}_{\nu \nu}(P)f(P)=f(P),
\end{equation}
and there is no momentum jump. An analogous procedure can be used to evaluate the higher order terms. For example, use of the second term in Eq.~(\ref{eq:Uexpansion}) in the memory kernel will yield contributions with products of three momentum jump operators. Typical contributions may be evaluated to give
\begin{eqnarray}
\hat{j}_{\nu \mu_1}(P) \hat{j}_{\mu_1 \mu_2 }(P) \hat{j}_{\mu_2 \nu }(P) f(P)&=&\hat{j}^{FS}_{\nu \nu}(P)f(P),\nonumber\\
\hat{j}_{\nu \beta}(P) \hat{j}_{\nu  \mu_1 }(P) \hat{j}_{\mu_1 \beta }(P) f(P)&=&\hat{j}^{FS}_{\nu \beta}(P)f(P).
\end{eqnarray}

\begin{figure}[htbp]
\begin{center}
\includegraphics[width=0.9\columnwidth]{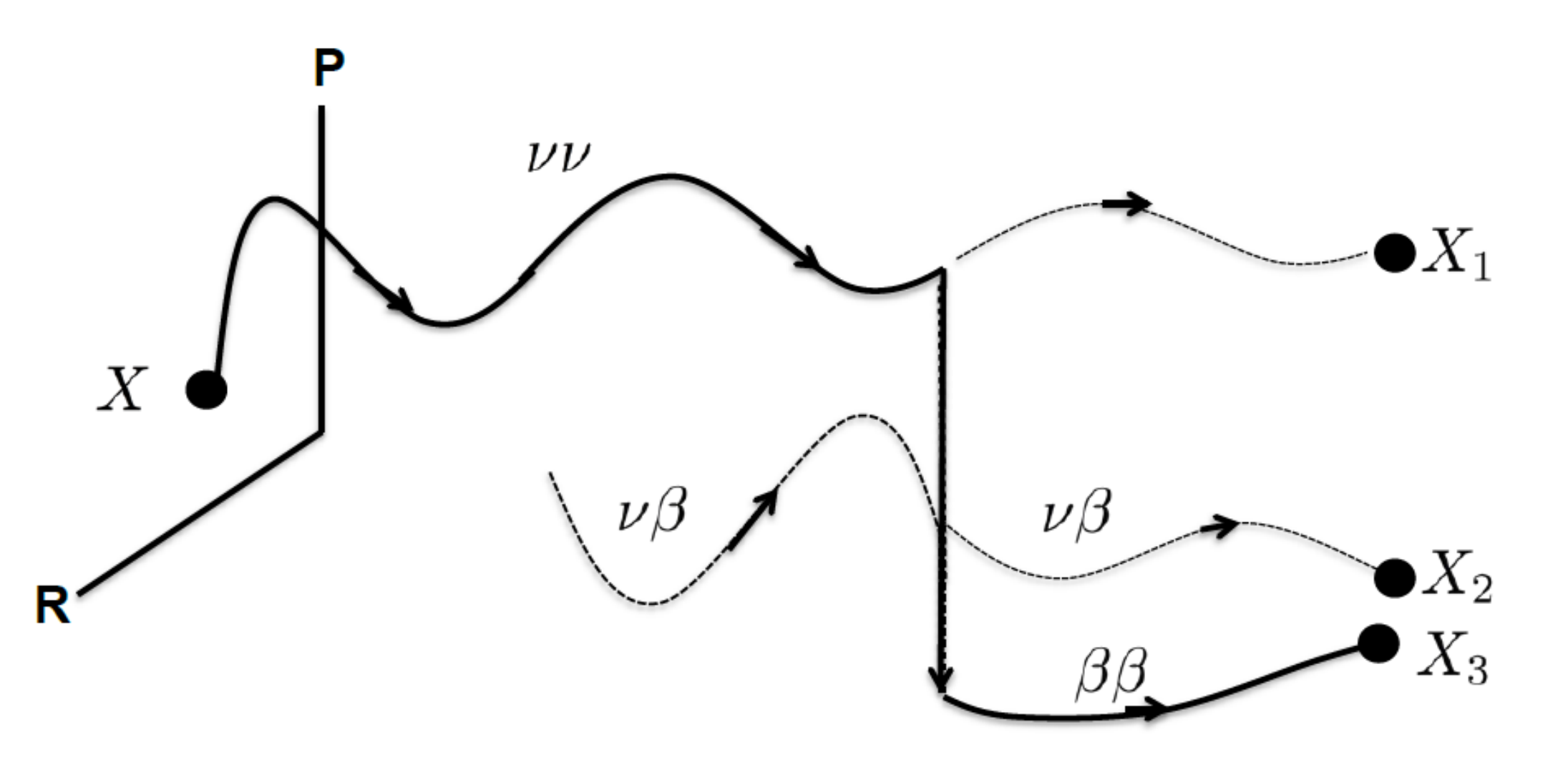}
%\vspace{20pt}
\caption{\label{fig:FS-traj} Fewest-switches-like stochastic trajectories corresponding to Eqs.~(\ref{eq:almost1}) and (\ref{eq:almost2}). The trajectory starts on the active $\nu$ surface (solid heavy line) . In the course of the evolution, as a result of nonadiabatic coupling contributions from other electronic density matrix elements arise (light doted lines)  although no nonadiabatic transitions have taken place and the system continues to evolve on the $\nu$ active surface. Later in the trajectory a nonadiabatic transition to the $\beta$ population state occurs as indicated by the heavy downward arrow.  Subsequently the $\beta$ state becomes the active surface and it is indicated by a solid heavy line. The trajectories are sketched assuming that the system is updated at times $\Delta t$ but this time interval may be taken to be infinitesimal. }
\end{center}
\end{figure}

Using these results the generalized master equation becomes
\begin{equation}\label{eq:master-nojump}
\frac{d}{d t} \rho^{\nu \nu}_W (X^\nu_t, t)= \int_0^t dt' \; \bar{{\mathcal M}}^\nu_{\nu \beta}(t,t') \hat{j}^{FS}_{\nu \beta}(X^\nu_{t'})\rho^{\beta \beta}_W (X^\nu_{t'}, t'),
\end{equation}
where the bar on $\bar{{\mathcal M}}^\nu$ is used to denote the fact that it no longer contains momentum jump operators. Having made these approximations we can return to the set of coupled equation that are equivalent to the generalized master equations,
\begin{equation}\label{eq:almost1}
\frac{d}{dt}{\rho}_W^{\nu \nu}(X^\nu_t,t) =-2 \Re \left(\frac{P^\nu_t}{M} \cdot d_{\nu \alpha'}(R^\nu_t)  \rho^{* \nu \alpha'}_W(X^\nu_t,t)\right ),
\end{equation}
and
\begin{eqnarray}\label{eq:almost2}
&& \hspace{-0.5cm} \frac{d}{d t} \rho^{\nu \alpha'}_W (X^\nu_t, t) = -i\omega_{\nu \alpha'}(R^\nu_t) \rho^{\nu \alpha'}_o (X^\nu_t,t) \\
&&-\sum_{\beta, (\beta \ne \nu, \alpha')} \left(\frac{P}{M} \cdot d_{\nu \beta}\rho^{\beta \alpha'}_W (X^\nu_t,t)
   +\frac{P}{M} \cdot d^*_{\alpha' \beta} \rho^{\nu \beta}_W (X^\nu_t,t)\right)\nonumber\\
&& -\frac{P}{M} \cdot d_{\nu \alpha'}\hat{j}^{FS}_{\nu \alpha'}(X^\nu_t) \rho^{\alpha' \alpha'}_W (X^\nu_t,t)
   -\frac{P}{M} \cdot d^*_{\alpha' \nu} \rho^{\nu \nu}_W (X^\nu_t,t). \nonumber
\end{eqnarray}
In writing Eq.~(\ref{eq:almost2}), in the last line  we have explicitly displayed the terms that couple the off-diagonal density matrix elements to diagonal elements to show where the fewest-switches momentum jump factors, $\hat{j}^{FS}_{\nu \alpha'}(X^\nu_t)$, appear.
With the exception of the momentum-jump operator in this equation, the pair of equations (Eqs.~(\ref{eq:almost1}) and (\ref{eq:almost2})) is identical to those that appear in FSSH (cf. Eqs.~(\ref{eq:sub-den-SH1}) and (\ref{eq:sub-den-SH2})). The trajectories that underlie these equations are indicated schematically in Fig.~\ref{fig:FS-traj}.

(3) To finish the story we must specify how these equations are to be solved by a stochastic algorithm. While the starting QCL equation treats all density matrix elements on a equal footing, the first two approximations leading to Eqs.~(\ref{eq:almost1}) and (\ref{eq:almost2}) served to give state $\nu$ a privileged status. This is the active surface on which the nuclear coordinates currently evolve. In addition to neglecting the decoherence terms, the approximations that specify the manner in which the momentum-jump operators act were made with the aim of considering transitions only between population states, so the stochastic algorithm should incorporate this feature. If the system is currently in state $\nu$, in the course of evolution on the $\nu$ surface the population can change at a rate given by Eq.~(\ref{eq:almost1}). Since other states are not currently active it seems appropriate to suppose that population changes involving this state arise solely from transitions out of the state. Transitions into state $\nu$ from other states would not be treated accurately since the nuclear evolution of those states is controlled by the active $\nu$ surface. In this context it seems reasonable to complete the final link to FSSH by choosing the transition rate in time interval $dt$ to be given by
\begin{equation}\label{eq:prob}
p_{\nu \to \beta}=\frac{2\Re(\frac{P}{M} \cdot d_{\nu \beta} \rho_W^{*\nu \beta})dt}{\rho_W^{\nu \nu}}\Theta\left(2\Re\left(\frac{P}{M} \cdot d_{\nu \beta} \rho_W^{*\nu \beta}\right)\right).
\end{equation}
where $\Theta(x)$ is a Heaviside function. While reasonable, there are aspects of this expression worth noting. The net rate in Eq.~(\ref{eq:almost1}) can take either sign and the Heaviside function in Eq.~(\ref{eq:prob}) restricts $p_{\nu \to \beta}$ to be positive. In contrast to the surface-hopping solution of the QCLE where reweighting factors enter the algorithm and the coupling terms can have either sign, this fewest-switches choice of transition rate is reasonable on physical grounds, given the form of the approximate equations of motion, and no reweighting of trajectories accompanies the nonadiabatic transitions. From these considerations, it is not obvious that modifications of the fewest-switches transition probability will improve the FSSH algorithm.

%%%%%%%%%%%%%%%%%%%%%%%%%%%%%
% Conclusion
%%%%%%%%%%%%%%%%%%%%%%%%%%%%%

\section{Discussion} \label{sec:con}

This is not the first time that connections between the QCLE and FSSH have been considered. As mentioned in the text, Subotnik, Ouyang and Landry~\cite{subotnik11c}, in an investigation with similar aims, constructed a nuclear-electronic density matrix starting with FSSH. In the course of the derivation a number of approximations and conditions had to be satisfied in order to obtain an evolution equation that was similar to but not exactly the same as the QCLE. Their derivation led to several ingredients that both justified some of the assumptions in FSSH and revealed some of its deficiencies. One of these major deficiencies was the lack a proper account of decoherence in the theory. The main decoherence factors they needed to append to the equations of motion are the same as those that enter in the treatment in this paper. In addition they showed how these decoherence factors could be approximated to yield tractable forms and how they are related to earlier suggestions for the treatment of decoherence.

The problem was approached from the opposite perspective in this study: the starting point was the QCLE and its solution by a surface-hopping algorithm. The QCLE was then transformed to a Lagrangian frame that moved with the dynamics on a specific adiabatic surface, and in this frame one could see what parts of the evolution operator needed to be modified to obtain FSSH. The resulting analysis does not constitute a derivation of FSSH since the result is obtained by discarding and approximating portions of the QCL operator, but it does provide considerable insight into the features that distinguish the quantum-classical Liouville and fewest-switches surface-hopping algorithms.

Several observations can be gleaned from the analysis presented in this paper. It is well known that the lack of a proper treatment of decoherence is one of the major shortcomings of FSSH and, as described in the text, various suggestion for how to incorporate decoherence in the surface-hopping framework have been proposed. Decoherence is taken into account in QCL dynamics and we have seen that the decoherence factor takes a suggestive form when this dynamics is viewed in a frame of reference that moves with the dynamics on a single active $\nu$ adiabatic surface. In the QCL dynamics the decoherence effects arise from transitions to and from the coherent evolution segments where the nuclear propagation occurs on the mean of two adiabatic surfaces and carries a phase. While the construction of decoherence factors in surface-hopping schemes often involves approximations whose validity is not fully determined, it is, in fact, very easy to simulate the evolution on the mean of two surfaces that describe the coherent (off-diagonal) evolution segments of the dynamics. So, to account for decoherence in surface hopping, rather than forcing the dynamics to evolve on single adiabatic surfaces, it is likely to be better to allow the system to jump to and propagate on both population and off-diagonal states.

As discussed earlier, this is the case for the QCL surface-hopping scheme where the evolution segments involve both diagonal and off-diagonal dynamics with transitions between them.  This is also the case for a recently-proposed surface-hopping scheme in Liouville space~\cite{prezhdo15}. That scheme incorporates transitions from diagonal to off-diagonal coherent evolution segments as in the surface-hopping solution of the QCLE, but the transition rates are approximated by forms analogous to those in FSSH. No reweighting is carried out and a prescription is given to obtain populations from the ensemble of trajectories.

Surface-hopping methods have considerable appeal when considering nonadiabatic dynamics since they provide a conceptually appealing way to view the dynamics. However, when one attempts to probe more deeply into their basis, the usual complexity of quantum mechanics, or even mixed quantum-classical mechanics, comes into play. The trajectories that comprise the ensemble that is used to compute observables are not independent and schemes must be devised to account for the correlations. This feature is manifest in the weights that the trajectories carry in the surface-hopping solution of the QCLE, as well as in other representation of this equation~\cite{kelly12}, and in a recent coherent state hopping method for nonadiabatic dynamics~\cite{martens15}.  Other research in this area has as its goal placing surface hopping on a more rigorous mathematical foundation.~\cite{lasser07,lu16} It seems that surface-hopping methods will continue to occupy our attention for some time.

%%%%
%%Acknowledgement
%%%%
%\begin{acknowledgments}
\section*{Acknowledgments}
This work was supported in part by a grant from the Natural Sciences and Engineering Council of Canada.

%\end{acknowledgments}

%\bibliographystyle{mybibfile}
%\bibliography{surfhopbib}
%\bibliography{surfhopbib}
%%  \bibliography{<your bibdatabase>}

\begin{thebibliography}{45}
\bibitem [{foo({\natexlab{a}})}]{footnote:nuclear-notation}%
  \BibitemOpen
  \href@noop {} {} ({\natexlab{a}}),\ \bibinfo {note} {More generally, we
  consider any open quantum system coupled to bath that can be described
  classically when isolated from the quantum system, but, for simplicity, in
  this paper we shall refer to the quantum and classical degrees of freedom as
  electronic and nuclear, respectively.}\BibitemShut {Stop}%
\bibitem [{foo({\natexlab{c}})}]{footnote:off-initial}%
  \BibitemOpen
  \href@noop {} {} ({\natexlab{c}}),\ \bibinfo {note} {In writing this equation
  we assumed that the off-diagonal elements are initialy zero. If this is not
  the case then an initial condition term will also be present in this
  equation. The analysis can be carried out when this term is
  included.}\BibitemShut {Stop}%
\bibitem [{foo({\natexlab{d}})}]{footnote:non-two-level}%
  \BibitemOpen
  \href@noop {} {} ({\natexlab{d}}),\ \bibinfo {note} {for a two-level system
  it takes a simpler form and is determined by the adiabatic
  frequencies.}\BibitemShut {Stop}%
\end{thebibliography}
%merlin.mbs aipnum4-1.bst 2010-07-25 4.21a (PWD, AO, DPC) hacked
%Control: key (0)
%Control: author (8) initials jnrlst
%Control: editor formatted (1) identically to author
%Control: production of article title (-1) disabled
%Control: page (0) single
%Control: year (1) truncated
%Control: production of eprint (0) enabled
%

\end{document}